\documentclass[superscriptaddress, letterpaper,twocolumn,prl,showpacs]{revtex4}
\usepackage{amsmath, amsthm, amssymb}
\usepackage[pdftex]{graphicx}
\usepackage{bm} 

\begin{document}
\title{Temperature dependence of the electronic structure and Fermi-surface reconstruction of Eu$_{1-x}$Gd$_{x}$O through the ferromagnetic metal-insulator transition}

\author{D.E. Shai}
\affiliation{Laboratory of Atomic and Solid State Physics, Department of Physics, Cornell University, Ithaca, New York 14853, USA}
\author{A.J. Melville}
\affiliation{Department of Materials Science and Engineering, Cornell University, Ithaca, New York 14853, USA}
\author{J.W. Harter}
\author{E.J. Monkman}
\author{D.W. Shen}
\affiliation{Laboratory of Atomic and Solid State Physics, Department of Physics, Cornell University, Ithaca, New York 14853, USA}
\author{A. Schmehl}
\affiliation{Zentrum f\"ur elektronische Korrelation und Magnetismus, Universit\"at Augsburg, Universit\"atsstra\ss e 1, 86159 Augsburg, Germany}
\author{D.G. Schlom}  
\affiliation{Department of Materials Science and Engineering, Cornell University, Ithaca, New York 14853, USA}
\affiliation{Kavli Institute at Cornell for Nanoscale Science, Ithaca, New York 14853, USA}
\author{K.M. Shen}
\email[Author to whom correspondence should be addressed: ]{kmshen@cornell.edu}
\affiliation{Laboratory of Atomic and Solid State Physics, Department of Physics, Cornell University, Ithaca, New York 14853, USA}
\affiliation{Kavli Institute at Cornell for Nanoscale Science, Ithaca, New York 14853, USA}

\begin{abstract}
We present angle-resolved photoemission spectroscopy of Eu$_{1-x}$Gd$_{x}$O through the ferromagnetic metal-insulator transition. In the ferromagnetic phase, we observe Fermi surface pockets at the Brillouin zone boundary, consistent with density functional theory, which predicts a half metal. Upon warming into the paramagnetic state, our results reveal a strong momentum-dependent evolution of the electronic structure, where the metallic states at the zone boundary are replaced by pseudogapped states at the Brillouin zone center due to the absence of magnetic long-range order of the Eu $4f$ moments.
\end{abstract}


\maketitle

EuO exhibits a remarkable array of magnetic properties which are induced by carrier doping, including colossal magnetoresistance ($\Delta \rho / \rho \approx 10^{6}$) \cite{shapira1973}, a large metal-insulator transition ($\Delta \rho / \rho \approx 10^{13}$) \cite{petrich1971, penney1972}, spin polarized carriers ($> 90 \%$) \cite{steeneken,schmehl2007}, and an enhancement of the Curie temperature ($T_c$) \cite{shafer1968}. Based on these properties and its compatibility with Si, GaN \cite{schmehl2007}, and GaAs \cite{swartz}, doped EuO has recently attracted attention to its potential in the development of spin valves and polarized injectors for spintronics \cite{santosPRB}. Despite nearly 40 years of research, a definitive picture of the momentum-resolved electronic structure of doped EuO across the metal-insulator transition is still lacking, in part due to the strong Coulomb repulsions in the half-filled Eu $4f$ shell which pose a challenge for calculations. Here we utilize angle-resolved photoemission spectroscopy (ARPES) to investigate epitaxial Eu$_{1-x}$Gd$_{x}$O thin films across the metal-insulator transition. Our measurements reveal a striking dichotomy of the behavior of doped carriers in momentum space which evolve between a delocalized, spin-polarized band at the Brillouin zone (BZ) boundary in the ferromagnetic (FM) metallic state, to localized, pseudogapped states at the BZ center in the paramagnetic (PM) state. 

35 nm thick Eu$_{1-x}$Gd$_{x}$O films were grown in both Veeco 930 and Veeco GEN10 oxide molecular-beam epitaxy (MBE) chambers on (110) terminated YAlO$_{3}$ substrates in adsorption-controlled conditions at a temperature of 350 $^{\circ}$C \cite{ross,sutartoGd}, where stoichiometric EuO can be produced without detectable concentrations of oxygen vacancies. The Eu flux was $1.1\times10^{14}$ atoms/cm$^{2}$s, and the Gd flux was varied to achieve different doping levels ($x$). Immediately following growth, the films were transferred to the ARPES chamber in less than 300 seconds under ultra-high vacuum ($2\times10^{-10}$ torr). ARPES measurements were performed using a VG Scienta R4000 spectrometer, with an instrumental energy resolution $\Delta E=25$ meV, He I$\alpha$ photons ($h\nu$ = 21.2 eV) and a base pressure typically better than $6\times10^{-11}$ torr. Film quality was monitored during growth using reflection high-energy electron diffraction (RHEED) and after ARPES measurements using low-energy electron diffraction (LEED), which shows a $1 \times 1$ surface structure. Results were confirmed by repeating measurements on over 35 individual samples and temperature-dependent measurements were confirmed by cycling samples from 140 K to 10 K and back to 140 K without noticeable degradation. Fermi surface (FS) maps were verified on multiple samples to check against the possibility of degradation. Additional \emph{ex situ} characterization was performed on films capped with amorphous Si using x-ray absorption spectroscopy at the SGM beamline at the Canadian Light Source to determine the true Gd concentration and x-ray diffraction to verify the film structure and phase purity \cite{supplemental}. 

\begin{figure}
	\includegraphics[width=1\columnwidth]{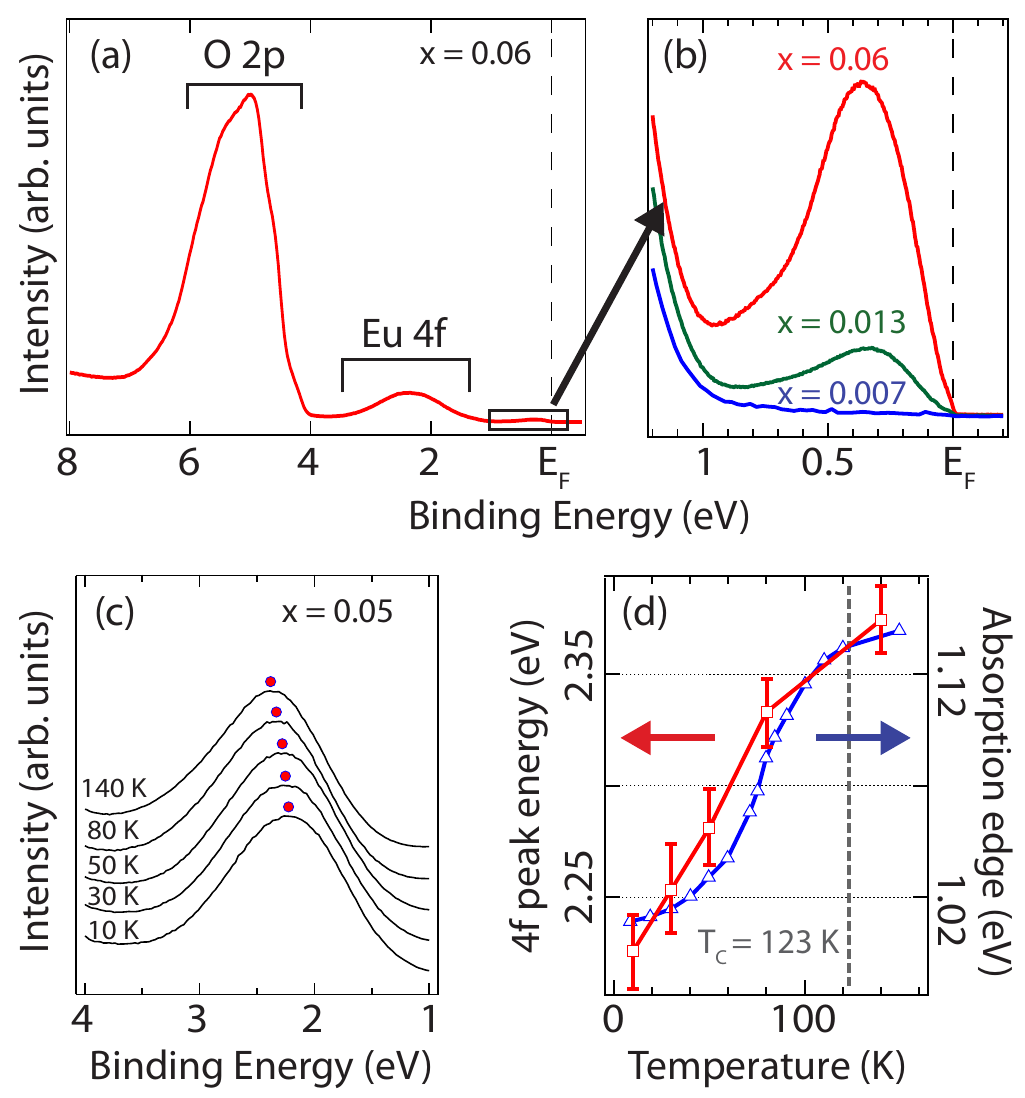}
	\caption{\label{fig:VB} (a)  Valence band of Eu$_{0.94}$Gd$_{0.06}$O at 10 K, integrated within a $k = 0\pm0.3$ \AA$^{-1}$ window along the line $k_{x}$ = $k_{y}$, showing the Eu 4$f$ states along with the O 2p states at higher binding energies.  (b) Near $E_{\rm F}$ states for Eu$_{1-x}$Gd$_x$O films at 10 K with $x = 0.007$, $0.013$ and $0.06$, integrated within a window of $\pm0.5$ \AA$^{-1}$. (c) Temperature dependence of the Eu 4$f$ band in Eu$_{0.95}$Gd$_{0.05}$O, showing a shift to lower binding energies with decreasing temperature. (d) Comparison of the $4f$ shift with the bulk redshift measured optically \cite{schoenes}.}
\end{figure}

In Fig. \ref{fig:VB}(a), we show the valence band for Eu$_{0.94}$Gd$_{0.06}$O with a nominal $T_c$ of 123 K, which consists of O $2p$ states between 4-6 eV and Eu $4f$ states around 1-3 eV binding energies, consistent with measurements on single crystals \cite{eastman} and undoped EuO films \cite{miyazaki}. In Fig. \ref{fig:VB}(b), we show the near $E_{\mathrm{F}}$ spectra for $x$ = 0.007, 0.013, and 0.06 doped samples at 10 K, exhibiting a monotonic increase in spectral weight with Gd (electron) doping. Films with $x < 0.007$ were highly insulating, did not exhibit near $E_{\mathrm{F}}$ spectral weight, and charged up electrostatically upon exposure to the photon beam; we could not observe evidence of metallic surface states predicted for undoped EuO \cite{schillerPRL}. Shown in Fig. \ref{fig:VB}(c) is the temperature dependence of the $4f$ peak in Eu$_{0.95}$Gd$_{0.05}$O around $(k_{x}, k_{y}) = (0, 0)$, which shifts to lower binding energies as the sample is cooled below $T_c$ (also reported by Miyazaki \emph{et al.} \cite{miyazaki}), while its lineshape remains largely unchanged. The temperature-dependent shift of the peak maximum is in close agreement with the optical redshift for a bulk sample of similar carrier concentration \cite{schoenes} shown in Fig. \ref{fig:VB}(d), indicating that our ARPES measurements are consistent with bulk properties measured using optical absorption. 

\begin{figure}
	\includegraphics[width=1\columnwidth]{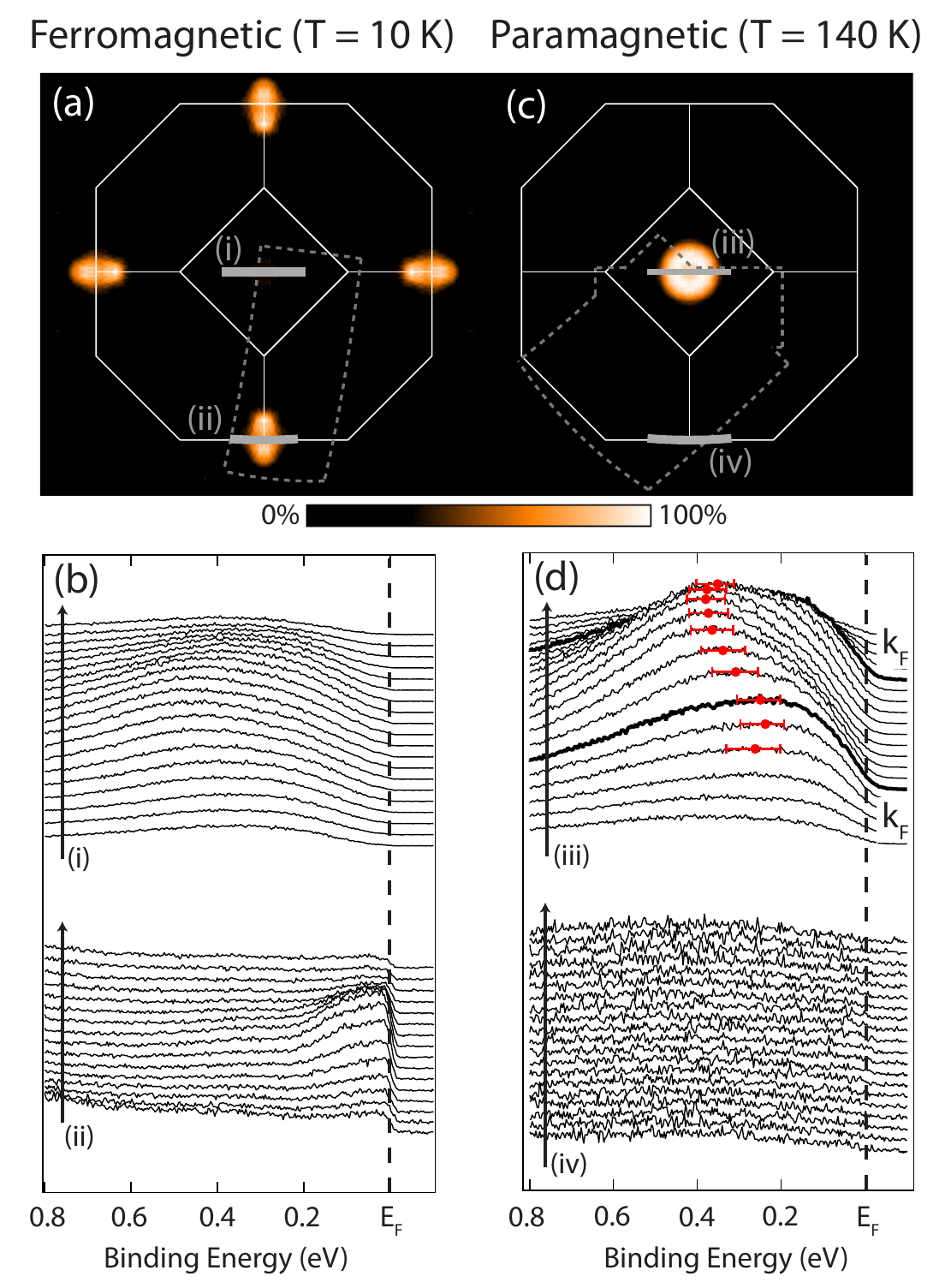}
	\caption{\label{fig:FSmap} Temperature-dependent near-$E_{\rm F}$ spectra for Eu$_{0.95}$Gd$_{0.05}$O. (a) Fermi surface map taken in the FM state at T = 10 K, with an integration window of $E_{\mathrm{F}}\pm$ 30 meV. The raw data were collected in the region within the dashed lines and symmetrized according to the 2D-projected BZ.   (b) Energy distribution curves (EDCs) taken at T = 10 K along cuts through the BZ boundary (0, 2$\pi/a$) and BZ center (0, 0).  The reciprocal space location of the cuts are indicated in (a).     (c) Fermi surface map taken above the Curie temperature at T = 140 K.  (d) EDCs at the BZ center and boundary taken at 140 K. The points for cut (iii) show the dispersion of the EDC peak  and $k_{\rm F}$ is determined by fitting the peaks in the momentum distribution curves at $E_{\rm F}$.  EDCs in (i) and (iii) are normalized to the 4f band peak intensity, and EDCs in (ii) and (iv) are normalized so the background at 1 eV matches (i) and (iii).}
\end{figure}

In order to address the nature of the spin-polarized metallic carriers in the FM state, we focus on the near $E_{\mathrm{F}}$ states at 10 K and will discuss the data above $T_{c}$ later in the text. A Fermi surface (FS) map is shown in Fig. \ref{fig:FSmap}(a).  The map clearly exhibits small elliptical pockets centered around the $X$ point ((0, 2$\pi/a$) in the 2D BZ) at the zone boundary. A perpendicular cut through this pocket in Fig. \ref{fig:FSmap}(b) reveals an electron-like band with a sharp Fermi cutoff at $E_{\mathrm{F}}$, indicating metallic states. Additionally, we also observe a low energy feature located near the ${\Gamma}$ point ((0, 0) in the 2D BZ), shown in Fig. \ref{fig:FSmap}(b). We attribute these to more deeply bound states (DBS) which are likely defect-derived and centered at a binding energy of 0.45 eV and exhibit only very weak dispersion and no appreciable spectral weight at $E_{\mathrm{F}}$, and therefore are not visible in the FS map. 

\begin{figure*}
	\includegraphics[width=1\textwidth]{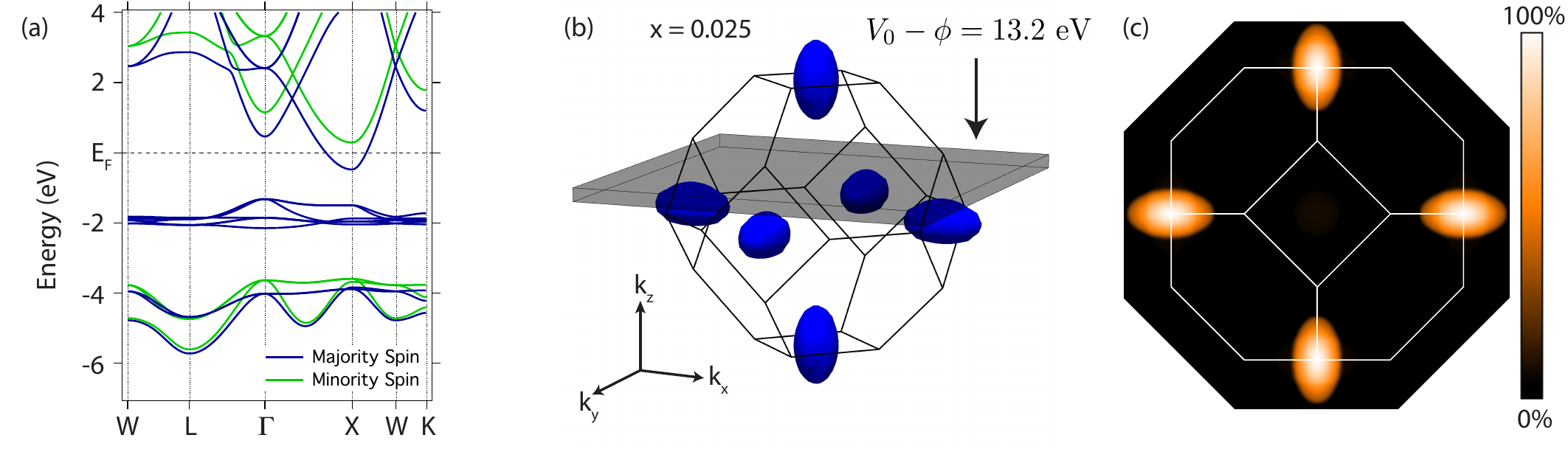}
	\caption{\label{fig:calculations} (a) Bandstructure for EuO along high symmetry directions in the fcc Brillouin zone. (b) Calculated spin-polarized Fermi surface for EuO, assuming a rigid shift of the Fermi level.  The grey box of width 1/$\lambda$, where $\lambda=8\ {\rm \AA}$ represents the integration window to account for the final state broadening \cite{supplemental}.  (c) Calculated ARPES Fermi surface, generated by averaging the data shown in (b) in the out-of-plane ($k_{z}$) direction (see text for details). }
\end{figure*}

In Fig. \ref{fig:calculations}, we compare our low temperature ARPES results to electronic structure calculations for Eu$_{1-x}$Gd$_{x}$O.  The calculations were performed using density functional theory (DFT) with the generalized gradient approximation plus on-site Coulomb interactions (GGA+U), based on the parameters determined by Ingle and Elfimov \cite{ilya, supplemental}.  In Fig. \ref{fig:calculations}(a), we plot the band structure along high symmetry lines, showing the minimum of the spin-polarized conduction band at the $X$ point. To treat the effect of carrier doping in the simplest manner possible, we perform a rigid shift of the chemical potential into the conduction band.  Fig. \ref{fig:calculations}(b) shows the calculated 3D FS consisting spin-polarized electron pockets centered around $X$.  In Fig. \ref{fig:calculations}(c), we present a simulated ARPES FS intensity map for $x = 0.025$ which takes into account the three-dimensionality of the FS using an inner potential of $V_{0}=13.2\ \rm{eV}+\phi$, where $\phi$ is the work function \cite{miyazaki}. We describe the treatment of the three-dimensionality of the electronic structure as well as ARPES measurements taken at a different photon energy (He II$\alpha$, $h\nu$ = 40.8 eV) in further detail elsewhere \cite{supplemental}, and will return to discuss our choice of the doping level later in the text.  The predicted ARPES FS is composed of  spin-polarized, elliptical electron pockets centered at (0, 2$\pi/a$) of primarily Eu $5d$ character. Although this is a simplified approach, the calculated FSes nevertheless appear to capture qualitatively the FS features experimentally observed in Fig. \ref{fig:FSmap}(a).

The qualitative agreement between the simulation and experiment demonstrates that DFT based approaches \cite{ilya, schillerSSC} can accurately treat the mobile carriers in the FM metallic state. This finding supports the numerous theoretical studies of EuO, including the possible effects of epitaxial strain on $T_c$ \cite{ilya}, realizing a new multiferroic from strained EuO \cite{bousquet}, and predictions for spin-polarized 2D electron gases at the interfaces of EuO-based superlattices \cite{Wang}. This agreement can also be taken as supporting evidence that the FM metal-insulator transition arises from an indirect exchange interaction between the Eu $5d$ conduction band and the FM ordered Eu $4f$ moments which lowers the bottom of the majority spin conduction band below $E_{\mathrm{F}}$. 

\begin{figure}[b]
	\includegraphics[width=1\columnwidth]{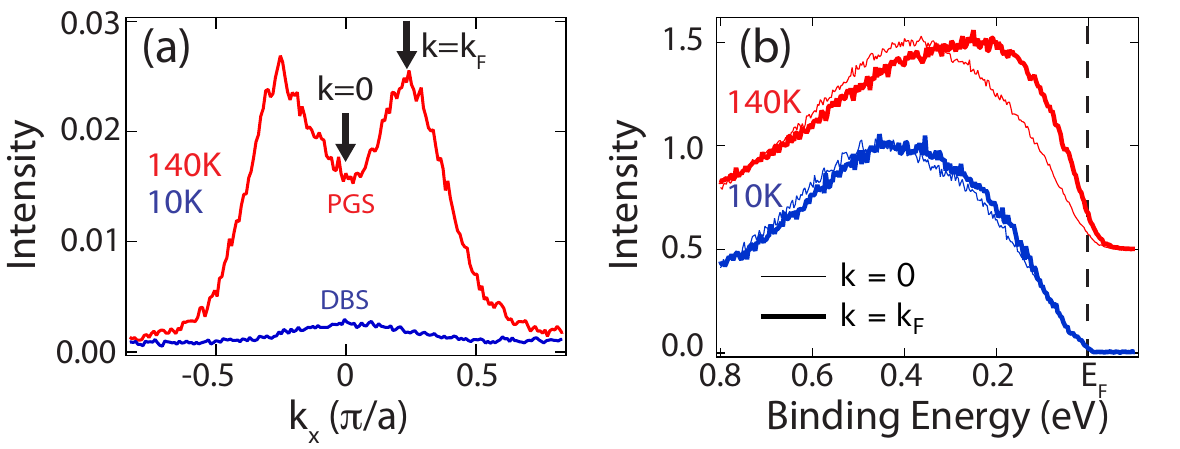}
	\caption{\label{fig:lineshape} 
	(a) Momentum distribution curves (MDCs) measured along ${k_{y}} = 0$ at 140 and at 10 K for Eu$_{0.95}$Gd$_{0.05}$O.  MDCs have been integrated within a $\pm$ 10 meV window about $\mathrm{E_{F}}$, and data has been normalized to the peak Eu 4f intensity. (b) Energy distribution curves taken at $k=0$ and $k=k_{F}$ (indicated in (a)) above and below $T_{c}$.  Spectra in (b) have been normalized to their peak intensity between 0.2-0.5 eV binding energy.}
\end{figure}

The more localized DBS states near (0, 0) are beyond the scope of our simple rigid band approximation. The DBS states exhibits some momentum dependence, indicating that these states comprise a defect band which is not completely localized in real space. Nevertheless, due to their vanishing spectral weight at $E_{\mathrm{F}}$, shown by the momentum distribution curve (MDC) in Fig. \ref{fig:lineshape}(a), the DBS likely play a negligible role in the low temperature conductivity, which is dominated by the metallic states at (0, 2$\pi/a$).

Upon warming into the PM insulating state (140 K), the FS map changes dramatically (Fig. \ref{fig:FSmap}(c)). Around (0, 2$\pi/a$), the metallic states are lifted above $E_{\mathrm{F}}$ (shown in Fig. \ref{fig:FSmap}(d) and the Supplemental Material \cite{supplemental}) due to the vanishing FM exchange splitting. Spectral weight is transferred to the BZ center, contributing to the ring of intensity around (0, 0), also shown in Fig \ref{fig:FSmap}(d). The contrast between the behavior of carriers at (0, 0) versus at (0, 2$\pi/a$) underscores the importance of employing a momentum-resolved probe in studying the properties of carrier doped EuO.

Our comparison of the lineshape at the BZ center below and above $T_{c}$ shown in Fig \ref{fig:lineshape}(b) and in the Supplemental Material \cite{supplemental} shows that the DBS (10 K) consist of broad and only weakly dispersive spectral weight.  In contrast, while the spectra above $T_{c}$ also exhibit the broad signature of the DBS at higher binding energy, they show an additional more highly dispersive component of spectral weight which sits closer to $E_{\rm F}$, indicating that another population of carriers has formed between the DBS and $E_{\rm F}$, which we call the pseudogapped states (PGS). The small but finite spectral weight of the PGS near $E_{\rm F}$ is likely responsible for the transport properties observed in the PM phase for films in this doping regime, which exhibit ``bad metal'' conduction at high temperatures, followed by a slight exponential upturn in the resistivity above $T_{c}$, indicating the formation of a small ($<$10 meV) activation gap \cite{mairoserDiplomaThesis,matsumoto}.  Such a gap is much smaller than our combined instrumental ($\Delta E=25$ meV) and thermal broadening at 140 K and therefore cannot be detected in our measurements.  This activated behavior in Eu$_{1-x}$Gd$_{x}$O is notably different from the variable range hopping conduction observed in the similar ferromagnetic metal-insulator compound Ga$_{1-x}$Mn$_{x}$As  \cite{sheu}.

Despite the increased intensity near $E_{\rm F}$, the PGS are still broad in energy and strongly suppressed in intensity within 200 meV of $E_{\rm F}$, forming a pseudogap reminiscent of features seen in the high-$T_c$ cuprates \cite{Shen1}, Fe$_{3}$O$_{4}$ \cite{schrupp}, and the manganites \cite{mannella, dessau}, all of which exhibit insulating behavior resulting from strong electron correlations.  For the case of Eu$_{1-x}$Gd$_{x}$O, such pseudogapped behavior could arise from incoherent carrier hopping due to an indirect exchange interaction between the dopant electrons and the randomly fluctuating background of Eu $4f^{7}$ spins above $T_{c}$.  The spectral weight transfer that we observe through the metallic transition is related to the transferring of electrons from the PGS to spin-polarized electron pockets at the BZ boundary.  This non-trivial behavior is unexpected and demonstrates that doped electrons in EuO do not simply enter non-dispersive donor impurity states.

Our observation of two populations of carriers above $T_{c}$ is consistent with the recent Hall measurements by Mairoser \emph{et al.} \cite{mairoser} which indicate that less than 50 \% of the Gd dopants are electrically active at 10 K.  The PGS electrons which are closer to $E_{\mathrm{F}}$ are electrically active above $T_{c}$ and are transferred to the EuO conduction band below $T_{c}$. On the other hand, the electrons comprising the DBS  well below $E_{\mathrm{F}}$ appear to remain inactive at all temperatures.  By integrating the spectral weight around (0, 0) from -1.0 eV to $E_{\mathrm{F}}$  (and assuming a spherical geometry) above and below $T_{c}$, we calculate that only 50 $\pm$ 10\% of the dopant electrons are transferred into the conduction band below $T_c$ \cite{supplemental}.  This incomplete ($\approx$ 50\%) carrier activation is why we have presented calculations for $x=0.025$ rather than the experimentally determined doping of $x=0.05$ in Fig. \ref{fig:calculations}.  Future work is needed to determine the origin of the segregation of carriers into active and inactive populations, but one possiblity is that the DBS represent the formation of a defect band resulting from Gd dopant clustering or Gd-induced oxygen vacancies.  We hope that our experimental observations will inspire more detailed experimental and theoretical studies of these bound states.

The behavior we have described in Eu$_{1-x}$Gd$_{x}$O exhibits striking similarities to the colossal magnetoresistive manganites. In both Eu$_{1-x}$Gd$_{x}$O and the manganites, sharp metallic features at $E_{\mathrm{F}}$ are clearly observed  by ARPES in the FM metallic state, but upon warming into the PM insulating state, only broad, dispersive, pseudogapped features remain \cite{mannella}. One profound difference between these systems is that in the manganites the pseudogapped bands in the PM state track the $\mathbf{k}$-space locations of the erstwhile FS in the FM state \cite{mannella, dessau}, while in Eu$_{1-x}$Gd$_{x}$O, the FM metallic and PM pseudogapped states exist in completely different regions of momentum space. This is somewhat unexpected, as dopants in n-type semiconductors would be expected to form below the conduction band minimum \cite{Efros}. This momentum-space dichotomy between the states at (0, 0) and those at (0, 2$\pi/a$) suggests that the DBS and PGS have extremely weak hybridization with the low-temperature metallic states and may explain the dramatic metal-insulator transition at low dopings.

\begin{figure}
	\includegraphics[width=1\columnwidth]{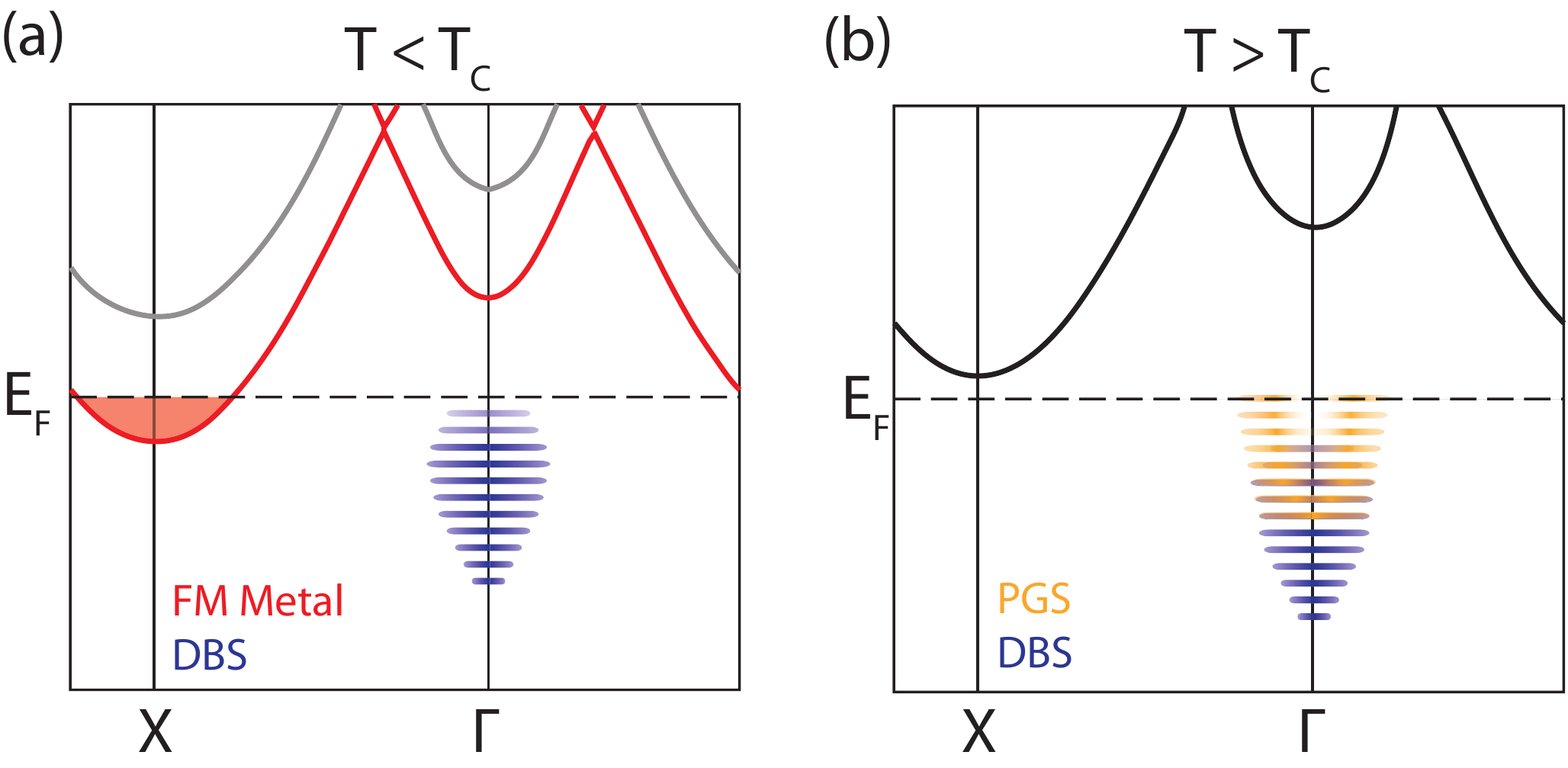}
	\caption{\label{fig:cartoon} (a)  Momentum space electronic structure in the ferromagnetic metallic phase and (b) in the paramagnetic semiconducting phase.  In (a), metallic carriers are indicated at $X$, while the DBS are shown at $\Gamma$. In (b) the PGS form at $\Gamma$ between the DBS and $E_{\rm F}$.}
\end{figure}

Our findings are summarized in the schematic in Fig. \ref{fig:cartoon}. Below $T_c$, metallic carriers are observed in the spin-polarized conduction band at (0, 2$\pi/a$) in addition to minimally dispersive deeply bound states at (0, 0).  Upon warming, the exchange splitting of the conduction band is reduced to zero and the metallic carriers and spectral weight at (0, 2$\pi/a$) are transferred to (0, 0),  where they populate a second dopant-induced state just below $E_{\rm F}$, the PGS.  These high temperature carriers exhibit a pseudogapped but dispersive lineshape, naturally explaining the increase in resistivity through the metal-insulator transition.  Our work reveals for the first time the nature of the FM metal-insulator transition in carrier doped EuO and uncovers an unusual momentum space dependence to the evolution of the electronic structure.  We hope that future experimental and theoretical works can resolve the origins of the momentum-space dichotomy and the DBS, since addressing these issues may be key to realizing the potential of EuO-based devices and applications.

We gratefully acknowledge I.S. Elfimov for assistance with DFT calculations; T.Z. Regier (SGM beam line at the Canadian Light Source) for his assistance with the XAS measurements; T. Mairoser, G.A. Sawatzky, and J. Geck for helpful discussions. Research was supported by the National Science Foundation through DMR-0847385 and the MRSEC program under DMR-1120296 (Cornell Center for Materials Research), and by the Research Corporation for Science Advancement (20025). D.E.S. acknowledges support from the National Science Foundation under Grant No. DGE-0707428 and NSF IGERT under Grant No. DGE-0654193.  AJM and DGS acknowledge the support of the AFOSR (FA9550-10-1-0123).


\begin{thebibliography}{99}

\bibitem{shapira1973} Y. Shapira, S. Foner, and T.B. Reed, Phys. Rev. B. {\bf 8,} 2299 (1973).

\bibitem{petrich1971} G. Petrich, S. von Moln\'ar, and T. Penney, Phys. Rev. Lett. {\bf 26,} 885 (1971).

\bibitem{penney1972} T. Penny, M.W. Shafer, and J.B. Torrance, Phys. Rev. B. {\bf 5} 3669 (1972).

\bibitem{schmehl2007} A. Schmehl {\it et al.}, Nature Mat. {\bf 6,} 882 (2007).

\bibitem{steeneken} P.G. Steeneken {\it et al.}, Phys. Rev. Lett. {\bf 88,} 047201 (2002).

\bibitem{shafer1968} M.W. Shafer and T.R. McGuire, J. Appl. Phys. {\bf 39,} 588 (1968).

\bibitem{swartz} A.G. Swartz {\it et al.}, Appl. Phys. Lett. {\bf 97,} 112509 (2010).

\bibitem{santosPRB} T.S. Santos and J.S. Moodera, Phys. Rev. B. {\bf 69}, 241203(R) (2004).

\bibitem{ross} R. Ulbricht {\it et al.}, Appl. Phys Lett. {\bf 93,} 102105 (2008).

\bibitem{sutartoGd} R. Sutarto {\it et al.}, Phys. Rev. B {\bf 80,} 085308 (2009).

\bibitem{supplemental} See Supplemental Material for discussion of the film characterization, additional photoemission data, and details for the supporting calculations.

\bibitem{eastman} D.E. Eastman, F. Holtzberg, and S. Methfessel, Phys. Rev. Lett. {\bf 23,} 226 (1969).

\bibitem{miyazaki} H. Miyazaki {\it et al.}, Phys. Rev. Lett. {\bf 102,} 227203 (2009).

\bibitem{schillerPRL} R. Schiller and W. Nolting, Phys. Rev. Lett. {\bf 86,} 3847 (2001).

\bibitem{schoenes} J. Schoenes and P. Wachter, Phys. Rev. B {\bf 9,} 3097 (1974).

\bibitem{ilya} N.J.C. Ingle and I.S. Elfimov, Phys. Rev. B {\bf 77,} 121202(R) (2008).

\bibitem{schillerSSC} R. Schiller and W. Nolting, Solid State Commun. {\bf 118,} 173 (2001).

\bibitem{bousquet} E. Bousquet, N.A. Spaldin, and P. Ghosez, Phys. Rev. Lett. {\bf 104}, 037601 (2010). 

\bibitem{Wang} Y. Wang {\it et al.}, Phys. Rev. B {\bf 79}, 212408 (2009).

\bibitem{mairoser} T. Mairoser {\it et al.}, Phys. Rev. Lett. {\bf 105,} 257206 (2010).

\bibitem{mairoserDiplomaThesis} T. Mairoser. Diploma thesis, Universit\"at Augsburg, 2009.

\bibitem{matsumoto} T. Matsumoto {\it et al.}, J. Phys.: Condens. Matter {\bf 16}, 6017 (2004)

\bibitem{sheu} B. Sheu {\it et al.}, Phys. Rev. Lett. {\bf 99,} 227205 (2007).

\bibitem{Shen1} K.M. Shen {\it et al.}, Phys. Rev. Lett. {\bf 93}, 267002 (2004). 

\bibitem{schrupp} D. Schrupp {\it et al.}, Europhys. Lett. {\bf 70}, 789 (2005). 

\bibitem{mannella} N. Mannella {\it et al.}, Nature {\bf 438}, 474 (2005). 

\bibitem{dessau} Y.D. Chuang {\it et al.}, Science {\bf 292}, 1509 (2001).

\bibitem{Efros} B.I. Shklovskii and A.L. Efros, {\it Electronic Properties of Doped Semiconductors}. (Springer-Verlag, Berlin, 1984).

\end{thebibliography}
\end{document}